\documentstyle{lamuphys}

\begin{document}

\title{Transformation of the Eilenberger Equations of\\
Superconductivity to a Scalar Riccati Equation}
\author{Nils Schopohl}
\institute{Eberhard-Karls-Universit\"{a}t T\"{u}bingen,
Institut f\"{u}r Theoretische Physik\\
Auf der Morgenstelle 14,
D-72076 T\"{u}bingen, Germany}
\titlerunning{Transformation of the Eilenberger Equations}
\maketitle

\begin{abstract}
A new parametrization of the Eilenberger equations of superconductivity in
terms of the solutions to a \emph{scalar} differential equation of the
Riccati type is introduced. It is shown that the quasiclassical propagator,
and in particular the local density of states, may be reconstructed, without
explicit knowledge of any eigenfunctions and eigenvalues, by solving a
simple \emph{initial value} problem for the linearized Bogoliubov-de Gennes-
equations. The Riccati parametrisation of the quasiclassical propagator
leads to a stable and fast numerical method to solve the Eilenberger
equations.
\end{abstract}

\section{Introduction}

According to the BCS theory of superconductivity the quasiparticle
excitations above the Cooper pairing groundstate depend on spin ( $\uparrow $
or $\downarrow $ ) and also on a particle-hole index ($+$ or $-$) which
indicates the flight direction of a quasiparticle (parallel or antiparallel
to the Fermi velocity ${\mathbf v}_F$ ). \emph{Coherent} superpositions of
such excitations form wave packets that transport energy, momentum, charge
and spin inside a superconductor.

In metals and alloys of interest to technical applications of
superconductivity the Cooper pairs display an \emph{even parity} symmetry
(spin singlet), and often the influence of paramagnetic effects ( Zeeman
splitting, Pauli limiting, spin-orbit coupling etc.) may be ignored. Then
spin and particle-hole indices may be identified. As a result the $4\times 4$%
-matrix equations of superconductivity may be simplified to $2\times 2$%
-matrix equations.

In the following we use notation such that ${\mathbf r}$ refers to a point in
position space (center of mass of a Cooper pair) , and ${\mathbf
p}_F=\hbar 
{\mathbf k}_F$ denotes a point on the Fermi surface $FS$ .

It is known that the characteristic length to heal a local (static)
perturbation of the Cooper pairing amplitude $\Delta ({\mathbf r},{\mathbf p}%
_{F})$ in a superconductor (due to the presence of an impurity, a vortex
line, an interface etc. ) is approximately $\xi =\frac{\hbar v_{F}}{\Delta
_{\infty }}$ , and often the quasiclassical condition $k_{F}\xi \gg 1$ is
fulfilled.

Then, as first shown by Eilenberger\cite{Eilenberger} and Larkin and
Ovchinnikov\cite{Larkin} , the relevant part of the physical information
coded in quantum mechanical expectation values ( for example the charge
density, the current, the pressure functional etc.) may be calculated more
efficiently with the help of the quasiclassical propagator 
\begin{equation}
\widehat{g}({\mathbf r};{\mathbf p}_{F},i\varepsilon _{n})=\left( 
\begin{array}{cc}
g({\mathbf r};{\mathbf p}_{F},i\varepsilon _{n}) & f({\mathbf r};{\mathbf p}%
_{F},i\varepsilon _{n}) \\ 
\bar{f}({\mathbf r};{\mathbf p}_{F},i\varepsilon _{n}) & \bar{g}({\mathbf r};%
{\mathbf p}_{F},i\varepsilon _{n})
\end{array}
\right)
\end{equation}
The quasiclassical propagator is, by definition, just the Green's function
of the Gorkov theory of superconductivity in a form where it has been
integrated with respect to the kinetic energy of the quasiparticles\cite
{Rainer}. Remarkably, $\widehat{g}({\mathbf r};{\mathbf p}_{F},i\varepsilon
_{n})$ may be also calculated directly solving a transport type system of 
\textit{ordinary } differential equations (the right hand side is a
commutator): 
\begin{eqnarray}
&&-i\hbar {\mathbf v}_{F}{\mathbf \cdot \nabla
\,}\widehat{g}({\mathbf r};{\mathbf %
p}_{F},i\varepsilon _{n})= \nonumber \\
&&\left[ \left( 
\begin{array}{cc}
i\varepsilon _{n}+{\mathbf v}_{F}\cdot \frac{e}{c}{\mathbf A}\left( {\mathbf r}%
\right) & -\Delta ({\mathbf r},{\mathbf p}_{F}) \\ 
\Delta ^{\dagger }({\mathbf r},{\mathbf p}_{F}) & -i\varepsilon _{n}-{\mathbf v}%
_{F}\cdot \frac{e}{c}{\mathbf A}\left( {\mathbf r}\right)
\end{array}
\right) ,\quad\widehat{g}({\mathbf r};{\mathbf p}_{F},i\varepsilon _{n})\right]
\label{qcl}
\end{eqnarray}
The physical solution to this equation must also fulfill a \emph{%
normalisation condition}: 
\begin{equation}
\widehat{g}({\mathbf r};{\mathbf p}_{F},i\varepsilon _{n})\cdot \widehat{g}(%
{\mathbf r};{\mathbf p}_{F},i\varepsilon _{n})=-\pi ^{2}\cdot \hat{1}
\label{normalization}
\end{equation}
General symmetries of the Gorkov Green's functions imply corresponding
symmetries of the quasiclassical propagator: 
\begin{eqnarray}
\bar{f}({\mathbf r};{\mathbf p}_{F},i\varepsilon _{n}) &=&-f\,^{*}({\mathbf r};%
{\mathbf p}_{F},-i\varepsilon _{n}) \\
\bar{g}({\mathbf r};{\mathbf p}_{F},i\varepsilon _{n}) &=&g({\mathbf r};-%
{\mathbf p}_{F},-i\varepsilon _{n}) \\
f({\mathbf r};-{\mathbf p}_{F},-i\varepsilon _{n}) &=&f({\mathbf r};{\mathbf p}%
_{F},i\varepsilon _{n}) \\
g({\mathbf r};{\mathbf p}_{F},i\varepsilon _{n}) &=&g^{*}({\mathbf r};{\mathbf p}%
_{F},-i\varepsilon _{n})
\end{eqnarray}
In equilibrium the quasiclassical propagator also displays a \emph{%
particle-hole} symmetry: 
\begin{equation}
\bar{g}({\mathbf r};{\mathbf p}_{F},i\varepsilon _{n})=-g({\mathbf r};{\mathbf p}%
_{F},i\varepsilon _{n})
\end{equation}
This means that the trace of $\widehat{g}({\mathbf r};{\mathbf p}%
_{F},i\varepsilon _{n})$ vanishes. As a traceless $2\times 2$-matrix the
square of $\widehat{g}$ should be equal to a multiple of unity: 
\begin{equation}
\widehat{g}({\mathbf r};{\mathbf p}_{F},i\varepsilon _{n})\cdot \widehat{g}(%
{\mathbf r};{\mathbf p}_{F},i\varepsilon _{n})=C\cdot \hat{1}
\end{equation}
Using the fact, that $\widehat{g}^{2}$ is a solution to the Eilenberger
equations (provided $\widehat{g}$ is a solution) , it follows that $-i\hbar 
{\mathbf v}_{F}{\mathbf \cdot \nabla }C=0$, i.e. the scalar $C$ is necessarily
a \emph{constant} along a straight line orientated parallel to the Fermi
velocity ${\mathbf v}_{F}$ . But $C$ could still be a function of the form $%
C=C({\mathbf r\wedge v}_{F};{\mathbf p}_{F},i\varepsilon _{n})$ . The
normalisation condition Eq.(\ref{normalization}) fixes $C$ such that $%
\widehat{g}^{2}=$ $-\pi ^{2}\cdot \hat{1}$ for \textit{all } straight lines
orientated parallel to ${\mathbf v}_{F}$ , and this for all Fermi momenta $%
{\mathbf p}_{F}$ on the Fermi surface and also for all Matsubara frequency $%
i\varepsilon _{n}$ . The particular value $C=-\pi ^{2}$ is chosen in order
to achieve consistency with the functional form of the quasiclassical
propagator in the bulk.

In thermal equilibrium the pair potential $\Delta ({\mathbf r},{\mathbf p}_{F})
$, the electrical current ${\mathbf J(r)}$ associated with a (stationary)
flow of quasiparticles, the local density of states $N({\mathbf r},E)$, the
Gibbs free energy $G_{S}$ of the superconducting state for weak coupling\cite
{Bardeen}, and other observables may be \emph{directly} calculated using the
quasiclassical propagator: 
\begin{eqnarray}
\Delta ({\mathbf r},{\mathbf p}_{F}) &=&\int_{FS}d{\mathbf p}_{F}^{\prime
}\;N_{FS}({\mathbf p}_{F}^{^{\prime }})V({\mathbf p}_{F},{\mathbf p}%
_{F}^{^{\prime }})\cdot k_{B}T\sum_{\left| \varepsilon _{n}\right| <\omega
_{c}}f({\mathbf r},{\mathbf p}_{F}^{\prime },i\varepsilon _{n}) \\
{\mathbf J(r)} &=&\frac{k_{B}}{\hbar }T\sum_{\varepsilon _{n}}\int_{FS}d%
{\mathbf p}_{F}^{\prime }\,N({\mathbf p}_{F}^{^{\prime }})\;{\mathbf v}%
_{F}^{\prime }\;g({\mathbf r},{\mathbf p}_{F}^{\prime },i\varepsilon _{n}) \\
N({\mathbf r},E) &=&-\frac{1}{\pi }\int_{FS}d{\mathbf p}_{F}^{\prime }\,N(%
{\mathbf p}_{F}^{^{\prime }})Im g({\mathbf r},{\mathbf p}_{F}^{\prime
},i\varepsilon _{n}\rightarrow E+i0^{+}) \\
&&  \nonumber \\
&& \\
G_{S}(T) &=&\int d{\mathbf r\;}\left[ 
\begin{array}{c}
-2T\cdot \int_{-\infty }^{\infty }dE\;N({\mathbf r},E)\cdot \ln \left( e^{%
\frac{E}{2T}}+e^{\frac{-E}{2T}}\right)  \\ 
\\ 
+\int_{FS}d{\mathbf p}_{F}\int_{FS}d{\mathbf p}_{F}^{\prime }\;\Delta
^{\dagger }({\mathbf r},{\mathbf p}_{F})\circ \left( V^{-1}\right) _{{\mathbf p}%
_{F}\,,\,{\mathbf p}_{F}^{^{\prime }}}\,\circ \Delta ({\mathbf r},{\mathbf p}%
_{F}^{\prime }) \\ 
\\ 
+\frac{1}{8\pi }\left( {\mathbf \nabla \wedge A(r)-B}_{ext}{\mathbf (r)}%
\right) ^{2}
\end{array}
\right]   \nonumber
\end{eqnarray}
In these expressions the function $N_{FS}({\mathbf p}_{F})$ denotes the (angle
resolved) density of states in the \emph{normal} phase at the Fermi level.
This function typically enters as a weight function into Fermi surface integrals
( $FS$ denotes the Fermi surface) of the Eilenberger propagator. In the isotropic
case $N_{FS}({\mathbf p}_{F})$ simplfies to the usual constant $N(0)$.

The calculation of Fermi surface integrals of the Eilenberger propagator becomes comparatively simple in
the \emph{bulk}, where the pair potential, $\Delta ({\mathbf p}_{F})$, is
independent on position ${\mathbf r}$, and where the quasiclassical propagator
assumes the form:

\begin{equation}
\widehat{g}({\mathbf p}_F,i\varepsilon _n)=\frac{-\pi }{\sqrt{\varepsilon
_n^2+\left| \Delta ({\mathbf p}_F)\right| ^2}}\cdot \left( 
\begin{array}{cc}
i\varepsilon _n & -\Delta ({\mathbf p}_F) \\ 
\Delta ({\mathbf p}_F)^{\dagger } & -i\varepsilon _n
\end{array}
\right)  \label{bulk}
\end{equation}

A considerably more complicated problem is posed when the pair potential
depends on position ${\mathbf r}$, for instance near a surface, in the
vicinity of an implanted impurity or ion, or around a flux line in a
type-II superconductor.

Usually the solution $\widehat{g}({\mathbf r};{\mathbf p}_{F},i\varepsilon
_{n})$ of the Eilenberger equations must be found numerically. But the task
is more difficult then just solving a differential equation. To determine
the pair potential $\Delta ({\mathbf r},{\mathbf p}_{F})$ and the magnetic
field ${\mathbf B(r)=}$ ${\mathbf \nabla \wedge A(r)}$ from the
(magnetostatic) Maxwell Equation , ${\mathbf \nabla \wedge B(r)}=\frac{4\pi }{%
c}{\mathbf J(r)}$, one needs to solve a (nonlinear) selfconsistency problem,
since ${\mathbf J(r)}$ and $\Delta ({\mathbf r},{\mathbf p}_{F})$ depend
themselves on $\widehat{g}({\mathbf r};{\mathbf p}_{F},i\varepsilon _{n})$.

\section{Eilenberger Equations along a Characteristic Line}

First we consider a layered material (normal axis parallel to ${\mathbf \hat{c%
}}$ ) assuming, for example, a Fermi velocity ${\mathbf v}_{F}$ that is
orientated predominantly within the $ab-$ plane (Fermi circle). Let the
triade $\left\{ {\mathbf \hat{a},\hat{b},\hat{c}}\right\} $ span an
orthonormal basis in the lab frame, while $\theta $ denotes the angle the
Fermi velocity ${\mathbf v}_{F}$ makes with the ${\mathbf \hat{a}}$-axis.
Clearly, along a straight line 
\begin{eqnarray}
{\mathbf r(}x) &=&{\mathbf \;}x{\mathbf \hat{v}+}y{\mathbf \hat{u}\;}
\label{straight} \\
&\equiv &{\mathbf \;}r_{a}(x)\widehat{{\mathbf a}}{\mathbf +}r_{b}(x)\widehat{%
{\mathbf b}} \\
-\infty &<&x<\infty  \nonumber
\end{eqnarray}
with ${\mathbf \hat{v}}$ and ${\mathbf \hat{u}}$ denoting unit vectors
(orientated parallel and orthogonal to ${\mathbf v}_{F}$ , respectively), 
\begin{eqnarray}
{\mathbf \hat{v}} &=&\cos (\theta ){\mathbf \hat{a}}+\sin (\theta ){\mathbf %
\hat{b}} \\
{\mathbf \hat{u}} &=&-\sin (\theta ){\mathbf \hat{a}}+\cos (\theta ){\mathbf %
\hat{b}}~,
\end{eqnarray}
the directional derivative ${\mathbf v}_{F}\cdot {\mathbf \nabla }$ ${\mathbf %
\,}$in the Eilenberger Equation Eq.(\ref{qcl}) is equivalent to an \emph{%
ordinary} derivative: 
\begin{equation}
\hbar {\mathbf v}_{F}\cdot {\mathbf \nabla \,}\widehat{g}({\mathbf r};{\mathbf p}%
_{F},i\varepsilon _{n})=\hbar v_{F}\frac{\partial }{\partial x}\widehat{g}%
\left[ {\mathbf r(}x{\mathbf )};{\mathbf p}_{F},i\varepsilon _{n}\right]
\end{equation}
The $\theta $-dependent parameter $y$ associated with such a \emph{%
characteristic} line ${\mathbf r}(x)$ (see Eq.(\ref{straight})) has
the natural meaning of an \textbf{impact} parameter. The straight line
 ${\mathbf r}(x)$ intersects with a fixed position point 
\begin{equation}
{\mathbf r}=r_{a}{\mathbf \hat{a}}+r_{b}{\mathbf \hat{b}}
\end{equation}
( there where the solution $\widehat{g}({\mathbf r};{\mathbf p}%
_{F},i\varepsilon _{n})$ is sought) at the particular parameter value $%
x=x_{P}$ . Introducing polar coordinates, 
\begin{equation}
r_{a}+ir_{b}=\sqrt{r_{a}^{2}+r_{b}^{2}}\,e^{i\phi }
\end{equation}
it is evident that 
\begin{equation}
r_{a}(x)+ir_{b}(x)=(x+iy)e^{i\theta }
\end{equation}
and 
\begin{equation}
x_{P}+iy=\sqrt{r_{a}^{2}+r_{b}^{2}}\,e^{i(\phi -\theta )}
\end{equation}

The extension to $3$-dimensions is
straightforward. For instance, for a spherical Fermi surface the unit vectors $\widehat{{\mathbf v}}$ and ${\mathbf \hat{u}%
}$ are parametrised by two angles, the azimutal angle $\theta \in \left[ 0,2\pi \right) $
and the polar angle $\chi \in \left[ 0,\pi \right) $, respectively: 
\begin{eqnarray}
\widehat{{\mathbf v}} &=&\sin \left( \chi \right) \left[ \cos (\theta )%
{\mathbf \hat{a}}+\sin (\theta ){\mathbf \hat{b}}\right] +\cos \left( \chi
\right) {\mathbf \hat{c}} \\
\widehat{{\mathbf u}} &=&\sin \left( \chi \right) \left[ -\sin (\theta )%
{\mathbf \hat{a}}+\cos (\theta ){\mathbf \hat{b}}\right] =\frac{\partial }{%
\partial \theta }\widehat{{\mathbf v}}
\end{eqnarray}
Again, along a straight line, ${\mathbf r}(x)=x\widehat{{\mathbf v}}+y{\mathbf %
\hat{u}+}z\widehat{{\mathbf v}}\wedge {\mathbf \hat{u}}$ , the directional
derivative in the Eilenberger equations becomes just an ordinary derivative.
Making the identification ${\mathbf r\equiv }r_{a}\widehat{{\mathbf a}}+r_{b}%
\widehat{{\mathbf b}}+r_{c}\widehat{{\mathbf c}}={\mathbf r}(x_{P})$ explicit
expressions for $x_{P}$ and both 'impact' parameters $y$ and $z$ in terms of 
$\theta $, $\chi $ and the cartesian coordinates $r_{a}$ , $r_{b}$ , $r_{c}$
of the fixed point ${\mathbf r}$ in position space are easily derived.

\smallskip Finally we simplify our notation by dropping the functional
dependence of $\Delta $ , ${\mathbf A}$ and $\stackrel{\wedge }{g}$ on
arguments that stay constant as $x$ varies from $-\infty $ to $\infty $ : 
\begin{eqnarray}
\Delta (x) &=&\Delta \left[ {\mathbf r(}x{\mathbf ),p}_F\right] \\
i\tilde{\varepsilon}_n(x) &=&i\varepsilon _n+{\mathbf v}_F\cdot \frac ec%
{\mathbf A}\left[ {\mathbf r(}x{\mathbf )}\right] \\
\hat{g}(x) &=&\widehat{g}\left[ {\mathbf r(}x{\mathbf )};{\mathbf p}%
_F,i\varepsilon _n\right]
\end{eqnarray}

\section{Riccati Parametrisation of Eilenberger Propagator}

Any traceless $2\times 2$-matrix may be expanded into the basis 
\begin{eqnarray}
\widehat{K}_{3} &=&\frac{1}{2}\widehat{\tau }_{3} \\
\widehat{K}_{\pm } &=&-\frac{i}{2}\cdot (\widehat{\tau }_{1}\pm i\widehat{%
\tau }_{2})
\end{eqnarray}
($\widehat{\tau }_{1}$, $\widehat{\tau }_{2}$ ,and $\widehat{\tau }_{3}$ are
standard $2\times 2$-Pauli matrices). We note that 
\begin{eqnarray}
\left[ \widehat{K}_{+},\widehat{K}_{-}\right] &=&-2\widehat{K}_{3} \\
\left[ \widehat{K}_{3},\widehat{K}_{\pm }\right] &=&\pm \widehat{K}_{\pm }
\end{eqnarray}
The Eilenberger equations may then be rewritten along a characteristic line $%
{\mathbf r}(x)$ orientated parallel to the Fermi velocity ${\mathbf v}_{F}$ in
the form:

\begin{equation}
\hbar v_F\frac \partial {\partial x}\widehat{g}(x)=\left[ -2\tilde{%
\varepsilon}_n(x)\widehat{K}_3+\Delta (x)\widehat{K}_{+}-\Delta ^{\dagger
}(x)\widehat{K}_{-}\;,\;\widehat{g}(x)\right]
\end{equation}

Let us consider the following $2\times 2$ system of \emph{ordinary}
differential equations for an auxiliary propagator $\widehat{Y}(x)$
(fundamental system): 
\begin{eqnarray}
\hbar v_F\frac \partial {\partial x}\widehat{Y}(x) &=&\left( -2\tilde{%
\varepsilon}_n(x)\widehat{K}_3+\Delta (x)\widehat{K}_{+}-\Delta ^{\dagger
}(x)\widehat{K}_{-}\right) \widehat{Y}(x) \\
\widehat{Y}(0) &=&\widehat{Y}_0
\end{eqnarray}
The initial values for $\widehat{Y}(x)$ at $x=0$ may be prescribed in terms
of a (yet unknown) constant $2\times 2$ matrix $\widehat{Y}_0$ of rank $2$ .
We may reconstruct the physical propagator $\widehat{g}$ , the one that
solves the Eilenberger equations \textit{and} respects the normalization
condition, $\widehat{g}(x)\cdot \widehat{g}(x)=-\pi ^2\cdot \hat{1}$ , from
the fundamental system $\widehat{Y}(x):$ 
\begin{equation}
\widehat{g}(x)=-\pi i\cdot \widehat{Y}(x)\cdot 2\widehat{K}_3\cdot \widehat{Y%
}^{-1}(x)  \label{physical}
\end{equation}
By putting $x$ at the end of the calculations to the particular value $x_P$
, the physical propagator (i.e. the input into the selfconsistency
equations) is recovered:

\begin{equation}
\widehat{g}(x_P)=\widehat{g}\left[ {\mathbf r(}x_P{\mathbf )};{\mathbf p}%
_F,i\varepsilon _n\right] \equiv \widehat{g}\left[ {\mathbf r};{\mathbf p}%
_F,i\varepsilon _n\right]
\end{equation}

The commutator in the Eilenberger equations implies the existence of several 
\emph{invariants} along the characteristic$\ $line ${\mathbf r}(x)$. For
example, if the normalization condition Eq.(\ref{normalization}) is
fulfilled at a particular fixed point ${\mathbf r}(x_{0})$, it will be
fulfilled everywhere along the line ${\mathbf r}(x)$. Likewise, the
determinant $\det \widehat{g}(x)$ and the trace $\mbox{tr}\widehat{g}(x)$
remain constant for $-\infty <x<\infty $.

Next we parametrize the $2\times 2$ matrix $\widehat{Y}(x)$ in the form 
\begin{equation}
\widehat{Y}=\exp (a_{+}\widehat{K}_{+})\exp (a_3\widehat{K}_3)\exp (a_{-}%
\widehat{K}_{-})
\end{equation}
in terms of three unknown functions $a_3(x)$ , $a_{+}(x)$ and $a_{-}(x)$ (
Euler like 'angles' in particle-hole space). The physical propagator, Eq.(%
\ref{physical}), assumes then the form 
\begin{equation}
\widehat{g}(x)=-\pi i\cdot \left[ 
\begin{array}{c}
\left[ 1-2a_{-}(x)a_{+}(x)\exp \left( -a_3(x)\right) \right] \cdot 2\widehat{%
K}_3\; \\ 
+\;a_{+}(x)\cdot \left[ a_{-}(x)a_{+}(x)\exp \left( -a_3(x)\right)
\;-1\right] \cdot 2\widehat{K}_{+}\; \\ 
+a_{-}(x)\cdot \exp \left[ -a_3(x)\right] \cdot 2\widehat{K}_{-}\;\;
\end{array}
\right]
\end{equation}
One finds from the differential equation for $\widehat{Y}(x)$ a set of three
coupled differential equations for $a_3(x)$ , $a_{+}(x)$ and $a_{-}(x)$ : 
\begin{eqnarray}
\stackrel{\cdot }{a}_3-2a_{+}\exp (-a_3)\stackrel{\cdot }{a}_{-} &=&-\frac{2%
\tilde{\varepsilon}_n}{\hbar v_F} \\
\exp (-a_3)\,\stackrel{\cdot }{a}_{-} &=&-\frac{\Delta ^{\dagger }}{\hbar v_F%
} \\
\stackrel{\cdot }{a}_{+}-a_{+}\stackrel{\cdot }{a}_3+a_{+}^2\exp (-a_3)%
\stackrel{\cdot }{a}_{-} &=&\frac \Delta {\hbar v_F}
\end{eqnarray}
Here $\stackrel{\cdot }{a}(x)\equiv \frac \partial {\partial x}a(x)$. It is
readily seen that the three equations decouple, and that $a_{-}$ and $a_3$
may be expressed in terms of $a_{+}$ only :

\begin{eqnarray}
a_3(x) &=&-\frac 2{\hbar v_F}\left[ \tilde{\varepsilon}_nx+\int_0^xds\Delta
^{\dagger }(s)a_{+}(s)\right] +a_3^{(0)}  \label{nesteda} \\
a_{_{-}}(x) &=&-\frac 1{\hbar v_F}\cdot \int_0^xds\,\Delta ^{\dagger
}(s)\exp \left[ a_3(s)\right] +a_{\_}^{(0)}
\end{eqnarray}
The differential equation that remains to be solved for $a_{+}(x)$ is a 
\emph{Riccati} equation: 
\begin{equation}
\hbar v_F\frac \partial {\partial x}a_{+}(x)+\left[ 2\tilde{\varepsilon}%
_n\,+\Delta ^{\dagger }(x)\,a_{+}(x)\right] \,a_{+}(x)-\Delta (x)=0
\label{riccatia}
\end{equation}
However, the accurate numerical calculation of the nested integral for $%
a_{-}(x)$ is time consuming (even on a fast computer). To overcome this
difficulty we use a trick.

Let $\widehat{g}_A(x)$ and $\widehat{g}_B(x)$ be two different solutions of
the Eilenberger equations. Then not only the linear combination $c_A\,%
\widehat{g}_A(x)+c_B\,\widehat{g}_B(x)$ is a solution , but the products $%
\widehat{g}_B(x)\cdot $ $\widehat{g}_A(x)$ and $\widehat{g}_A(x)\cdot $ $%
\widehat{g}_B(x)$ are solutions as well. For example, the linear combination 
$\widehat{g}_B(x)\cdot \widehat{g}_A(x)-\widehat{g}_A(x)\cdot \widehat{g}%
_B(x)$ solves the Eilenberger equations \textit{and } fulfills the necessary
condition $\mbox{tr}\widehat{g}(x)=0$ .

Let us construct two particular \emph{zero} trace solutions to the
Eilenberger equations: 
\begin{eqnarray}
\widehat{g}_{A}(x) &=&\widehat{Y}_{A}(x)\cdot \widehat{K}_{-}\;\cdot \left[ 
\widehat{Y}_{A}(x)\right] ^{-1}\   \label{gA} \\
\widehat{g}_{B}(x) &=&\widehat{Y}_{B}(x)\cdot \widehat{K}_{+}\cdot \left[ 
\widehat{Y}_{B}(x)\right] ^{-1}  \label{gB}
\end{eqnarray}
with 
\begin{eqnarray}
\widehat{Y}_{A} &=&\exp (a_{+}\widehat{K}_{+})\exp (a_{3}\widehat{K}%
_{3})\exp (a_{-}\widehat{K}_{-}) \\
\widehat{Y}_{B} &=&\exp (b_{-}\widehat{K}_{-})\exp (b_{3}\widehat{K}%
_{3})\exp (b_{+}\widehat{K}_{+})
\end{eqnarray}
denoting two equivalent fundamental systems $\widehat{Y}_{A}(x)$and $%
\widehat{Y}_{B}(x)$. The different order of factors in the defining
expressions for $\widehat{Y}_{A}$ and $\widehat{Y}_{B}$ serves the purpose
to avoid the difficult terms $a_{-}(x)$ and $b_{+}(x)$ in the expressions
for $\widehat{g}_{A}$ and $\widehat{g}_{B}$ . The evaluation of nested
integrals, see Eq.(\ref{nesteda}), is then not necessary.

The set of equations fulfilled by $b_3(x)$ and $b_{\pm }(x)$ is only
slightly different from the one for $a_3(x)$ and $a_{\pm }(x)$ : 
\begin{eqnarray}
\stackrel{\cdot }{b}_3+2b_{-}\exp (b_3)\stackrel{\cdot }{b}_{+} &=&-\frac{2%
\tilde{\varepsilon}_n}{\hbar v_F} \\
\exp (b_3)\,\stackrel{\cdot }{b}_{+} &=&\frac \Delta {\hbar v_F} \\
\stackrel{\cdot }{b}_{-}+b_{-}\stackrel{\cdot }{b}_3+b_{-}^2\exp (b_3)%
\stackrel{\cdot }{b}_{+} &=&-\frac{\Delta ^{\dagger }}{\hbar v_F}
\end{eqnarray}
Here $\stackrel{\cdot }{b}(x)\equiv \frac \partial {\partial x}b(x)$. It is
readily seen that the three equations decouple, and that $b_{+}(x)$ and $%
b_3(x)$ may be expressed in terms of $b_{-}(x)$ only :

\begin{eqnarray}
b_3(x) &=&-\frac 2{\hbar v_F}\left[ \tilde{\varepsilon}_nx+\int_0^xds\,%
\Delta (s)b_{-}(s)\right] +b_3^{(0)}  \label{nestedb} \\
b_{+}(x) &=&\frac 1{\hbar v_F}\int_0^xds\,\Delta (s)\exp \left[
-b_3(s)\right] +b_{+}^{(0)}
\end{eqnarray}
The differential equation to be solved for $b_{-}(x)$ is also a Riccati
equation: 
\begin{equation}
\hbar v_F\frac \partial {\partial x}b_{-}(x)-\left[ 2\tilde{\varepsilon}%
_n+\Delta (x)\,b_{-}(x)\right] \,b_{-}(x)+\Delta ^{\dagger }(x)=0
\label{riccatib}
\end{equation}

We observe that any solution of this differential equation is related to the
Riccati equation Eq.(\ref{riccatib}) via a reciprocity relation:

\textit{\ If \ }$a_{+}(x)$\textit{\ \ solves Eq.(\ref{riccatia}), then} 
\begin{equation}
\mathit{\ }b_{-}(x)=-\frac 1{a_{+}(x)}
\end{equation}
\textit{\ \ solves Eq.(\ref{riccatib}).}

\smallskip
We continue now our construction of the physical propagator. From the
defining equations, Eq.(\ref{gA}) and Eq.(\ref{gB}), we find the following
explicit expressions: 
\begin{eqnarray}
\widehat{g}_{A} &=&\exp (-a_{3})\left( \widehat{K}_{-}-2a_{+}\,\widehat{K}%
_{3}+a_{+}^{2}\,\widehat{K}_{+}\right)  \label{gAx} \\
\widehat{g}_{B} &=&\exp (b_{3})\left( \widehat{K}_{+}+2b_{-}\,\widehat{K}%
_{3}+b_{-}^{2}\,\widehat{K}_{-}\right)  \label{gBx}
\end{eqnarray}
Note that the square of $\widehat{g}_{A}(x)$ and $\widehat{g}_{B}(x)$
vanishes identically, 
\begin{equation}
\widehat{g}_{A}\cdot \widehat{g}_{A}\,=\widehat{0}=\,\widehat{g}_{B}\cdot 
\widehat{g}_{B}  \label{gagb}~,
\end{equation}
because $\widehat{K}_{\pm }^{2}\equiv 0$ . For $x\rightarrow \pm \infty $
the propagators $\widehat{g}_{A}$ and $\,\widehat{g}_{B}$ 'explode', i.e. 
\begin{equation}
\widehat{g}_{A,B}\sim \exp (\pm \frac{2x}{\hbar v_{F}}\sqrt{\tilde{%
\varepsilon}_{n}^{2}+\left| \Delta \right| ^{2}})
\end{equation}
On the other hand, the commutator $\left[ \widehat{g}_{A}(x),\widehat{g}%
_{B}(x)\right] $ remains bounded in the limit $x\rightarrow \pm \infty $.
The observation that a \emph{bounded} solution to the Eilenberger equations
may be constructed using the commutator of two \emph{unbounded} solutions $%
\widehat{g}_{A}(x)$ and $\widehat{g}_{B}(x)$ is the well known 'explosion'
trick \cite{Thuneberg}.

The \emph{general} (particle-hole symmetric) solution to the Eilenberger
equations (\ref{qcl}) may be given in the form 
\[
\widehat{g}(x)=c_{A}\widehat{g}_{A}(x)+c_{B}\widehat{g}_{B}(x)+[\widehat{g}%
_{A}(x),\ \widehat{g}_{B}(x)] 
\]
Here, $c_{A}$ and $c_{B}$ represent initial values, $b_{3}(0)$ and $a_{3}(0)$
, to the functions $b_{3}(x)$ and $a_{3}(x)$. Of course, in an unbounded
region exploding solutions must be forbidden. Then the physical propagator $%
\widehat{g}$ must be written, on either side of the turning point $x=0$ , as
a superposition of a \emph{decaying} solution and a \emph{bounded} solution: 
\begin{equation}
\widehat{g}(x)=\left\{ 
\begin{array}{c}
c_{B}\widehat{g}_{B}(x)+[\widehat{g}_{A}(x),\ \widehat{g}_{B}(x)]\;\;
{\rm if\ } x>0 \\ 
c_{A}\widehat{g}_{A}(x)+[\widehat{g}_{A}(x),\ \widehat{g}_{B}(x)]\;\;{\rm
if\ }x<0
\end{array}
\right. \   \label{gx}
\end{equation}
The square of $\widehat{g}(x)$ is in this case \emph{independent} on the
constants $c_{A}$ and $c_{B}$ : 
\begin{equation}
\widehat{g}(x)\,\cdot \widehat{g}(x)=[\widehat{g}_{A}(x),\ \widehat{g}%
_{B}(x)]\cdot [\widehat{g}_{A}(x),\ \widehat{g}_{B}(x)]=-\pi ^{2}\cdot \hat{1%
}
\end{equation}
It is not difficult to show, that $c_{A}=0=c_{B}$ , provided the propagators 
$\widehat{g}_{A}(x)$ and $\widehat{g}_{B}(x)$ are \emph{continuous} at $x=0$.

In fact, let us assume the contrary: $c_A\cdot $ $c_B$ $\neq 0$ . Continuity
of $\widehat{g}(x)$ at $x=0$ leads to 
\begin{equation}
c_B\widehat{g}_B(0^{+})+[\widehat{g}_A(0^{+}),\ \widehat{g}_B(0^{+})]=\hat{g}%
(0)=c_A\widehat{g}_A(0^{-})+[\widehat{g}_A(0^{-}),\ \widehat{g}_B(0^{-})]
\label{continuity}
\end{equation}
Both solutions, $\widehat{g}_A(x)$ and $\widehat{g}_B(x)$ , are continuous
at $x=0$ \textbf{. }Then it follows from Eq.(\ref{continuity}) : $c_B%
\widehat{g}_B(0)=c_A\widehat{g}_A(0)$. This implies, in turn, a vanishing
commutator, $[\widehat{g}_A(0),\ \widehat{g}_B(0)]=0$, since $\widehat{g}%
_B(0)$ and $\widehat{g}_A(0)$ become proportional. Also, the \emph{physical}
solution $\hat{g}(0)$ at $x=0$ must fulfill the normalization condition,i.e. 
$\hat{g}(0)\cdot \hat{g}(0)=-\pi ^2\cdot \hat{1}$ . But $\widehat{g}%
_B(0)\cdot \widehat{g}_B(0)=\hat{0}=$ $\widehat{g}_A(0)\cdot \widehat{g}%
_A(0) $ according to Eqs.(\ref{gagb}). This is a contradiction! Hence $%
c_A=0=c_B$ .

The conclusion is that in an \emph{infinitely extended} system the physical propagator 
$\widehat{g}(x)$ is completely determined by the commutator of the
'exploding' solutions:

\begin{equation}
\widehat{g}(x)=\left[ \widehat{g}_{A}(x),\widehat{g}_{B}(x)\right] =\exp
(b_{3}-a_{3})\cdot \left[ 
\begin{array}{cc}
1-\left( a_{+}b_{-}\right) ^{2} & 2ia_{+}\left( 1+a_{+}b_{-}\right) \\ 
-2ib_{-}\left( 1+a_{+}b_{-}\right) & -1+\left( a_{+}b_{-}\right) ^{2}
\end{array}
\right]
\end{equation}
Next we check the normalisation condition: 
\begin{eqnarray}
\widehat{g}(x)\,\cdot \widehat{g}(x) &=&[\widehat{g}_{A}(x),\ \widehat{g}%
_{B}(x)]\cdot [\widehat{g}_{A}(x),\ \widehat{g}_{B}(x)] \\
&=&\left[ g_{3}(x)\cdot g_{3}(x)+g_{+}(x)\cdot g_{-}(x)\right] \cdot 
\widehat{1}  \nonumber \\
&=&\left[ 1+a_{+}(x)b_{-}(x)\right] ^{4}\cdot \exp \left[
2b_{3}(x)-2a_{3}(x)\right] \cdot \widehat{1}  \nonumber \\
&=&C\cdot \hat{1}  \nonumber
\end{eqnarray}
Indeed, $C$ is a \emph{constant} multiple of unity: 
\begin{equation}
\frac{\partial }{\partial x}\left\{ \left[ 1+a_{+}(x)b_{-}(x)\right]
^{4}\cdot \exp \left[ 2b_{3}(x)-2a_{3}(x)\right] \right\} =0
\end{equation}
From the normalisation condition, $C=-\pi ^{2}$ , there follows for all $x$
(up to a sign $\pm $ that is chosen to coincide with the bulk propagator): 
\begin{equation}
\exp \left[ b_{3}(x)-a_{3}(x)\right] =\frac{-\pi i}{\left[ 1+a_{+}(x)b_{-}(x)\right] ^{2}}
\end{equation}
Then the Eilenberger propagator may be parametrised
in the form:

\begin{equation}
\widehat{g}(x)\,=\frac{-\pi i}{%
1+a(x)\cdot b(x)}\cdot \left[ 
\begin{array}{cc}
1-a(x)\cdot b(x) & 2i\cdot a(x) \\ 
-2i\cdot b(x) & -1+a(x)\cdot b(x)
\end{array}
\right]  \label{riccpar}
\end{equation}

Here and in the following we use notation such that $b_{-}(x)$ $\equiv b(x)$
and $a_{+}(x)\equiv a(x)$, since the other functions $a_{-}(x)$,$a_{3}(x)$ and $b_{+}(x)$,$b_{3}(x)$
are obsolete for the parametrisation of the Eilenberger propagator.
It is remarkable that the solution to the Eilenberger equations (\ref
{qcl}) may be given a representation where it depends just on the solution
of an \emph{initial value problem} to a \emph{scalar} differential equation of the 
Riccati type\cite{schopohl2},\cite{schopohl and maki}.

To integrate the Riccati equations (\ref{riccatia}, \ref{riccatib}) in a \emph{stable} manner we need 
suitable initial values for the functions $b(x)$ and $a(x)$. 
For  $i\varepsilon _{n}$ situated in the \emph{upper half} of the complex plane the function $a(x)$ may be found in
a \emph{stable} manner integrating Eq.(\ref{riccatia}) as an initial value problem from $x=-\infty $
towards \emph{increasing} $x$-values, while the function $b(x)$ may be found
integrating Eq.(\ref{riccatib}) as an initial value problem from $x=+\infty $ \emph{backwards }%
towards \emph{decreasing} $x$-values. The initial values for
 $a(x)$ at $x=-\infty $ and $b(x)$ at $x=+\infty $ are 

\begin{eqnarray}
a(-\infty ) &=&\frac{\Delta (-\infty )}{\varepsilon _{n}+\sqrt{\varepsilon_{n}^{2}+\left| \Delta (-\infty )\right|^{2}}} \label{a(-00)} \\
b(+\infty ) &=&\frac{\Delta ^{\dagger }(+\infty )}{\varepsilon _{n}+\sqrt{\varepsilon _{n}^{2}+\left| \Delta (+\infty )\right| ^{2}}}\label{b(+00)}
\end{eqnarray} 
provided $i\varepsilon _{n}$ is in the \emph{upper half} of the complex plane. 

The differential equations to be solved are:
\begin{eqnarray}
\hbar v_{F}\frac{\partial }{\partial x}a(x)+\left[ 2\tilde{\varepsilon}%
_{n}\,+\Delta ^{\dagger }(x)\,\cdot a(x)\right] \,\cdot a(x)-\Delta (x) &=&0
\label{a(x)} \\
\hbar v_{F}\frac{\partial }{\partial x}b(x)-\left[ 2\tilde{\varepsilon}%
_{n}+\Delta (x)\,\cdot b(x)\right] \cdot \,b(x)+\Delta ^{\dagger }(x) &=&0
\label{b(x)}
\end{eqnarray}

Sometimes knowledge of just one of the functions, say $a(x)$, along a $\ $line ${\mathbf r(}x%
{\mathbf )}$ (for $-\infty <x<\infty $) suffices to fix the other function, $b(x)$, along the same line.
 An illustrative example is provided by a single
 \emph{cylindrically symmetric} vortex line, orientated parallel to  ${\mathbf \hat{c}}$, and 
centered at the origin of the $ab$ -plane, say at ${\mathbf R}={\mathbf 0}$. 
Due to energetic reasons, of course, only a single quantum of circulation, $\frac{h}{2m}$, is attached to the vortex.
The corresponding pair potential becomes along the straight line 
 ${\mathbf r}(x)=r_{a}(x){\mathbf \hat{a}} + r_{b}(x){\mathbf \hat{b}}$ a function of $x$ (and also of the impact parameter $y$)
 of the form:
\[
\Delta ({\mathbf r}(x),{\mathbf p}_{F})=F(\sqrt{x^{2}+y^{2}},\theta )\cdot 
\frac{x+iy}{\sqrt{x^{2}+y^{2}}}\cdot e^{i\theta } 
\]
The prefactor $F(\sqrt{x^{2}+y^{2}},\theta )$ is a suitable 'form factor' to shape the vortex core.
We see from Eqs.(\ref{a(x)},\ref{b(x)}) that in the presence of such a vortex line, $b(x)$ is related to $a(x)$ by symmetry:
\begin{equation}
b(x)=-a(-x)e^{-2i\theta }
\end{equation}
Using Eq.(\ref{riccpar}) it follows that the corresponding Eilenberger propagator $\widehat{g}$ 
for negative $x$ is related to the propagator for positive $x$ by the relation: 
\begin{equation}
\widehat{g}(-x)=-e^{i\theta \hat{\tau}_3}\cdot \hat{\tau}_2\cdot \widehat{g}%
(x)\cdot \hat{\tau}_2\cdot e^{i\theta \hat{\tau}_3}
\end{equation}

\smallskip

To determine the local density of states one needs the retarded and the advanced propagator of the quasiclassical theory.
Actually, in equilibrium, only the retarded (or advanced) propagator is needed in the calculations, since both propagators
are related to each other by complex conjugation. 
A convenient numerical method for the calculation of the retarded propagator 
 $\hat{g}^{(ret)}\left( {\mathbf r},\theta ,E\right) $
is to replace the discrete Matsubara frequency $i\varepsilon _{n}$ 
according to the prescription $i\varepsilon _{n}\rightarrow E+i0^{+}$,
 and to solve the Riccati equations, Eqs.(\ref{a(x)},\ref{b(x)}), as functions
 of the energy $E$  and of the impact parameters $y$ and $z$.

If the denominator of the quasiclassical propagator, $1+a(x)\cdot
b(x)$, becomes equal to zero at a point ${\mathbf r}(x_{0})$ for a
characteristic energy $E=E_{b}$, it vanishes indeed for 
\emph{all} $x$ along the trajectory ${\mathbf r}(x)$: 
\begin{equation}
\left[ 1+a(x)\cdot b(x)\right] =\exp \left[ \frac{1}{\hbar v_{F}}%
\int_{x_{0}}^{x}ds\;\left( \Delta (s)\,b(s)-\Delta ^{\dagger
}(s)\,a(s)\right) \right] \cdot \left[ 1+a(x_{0})\cdot
b(x_{0})\right]  \label{bound state condition}
\end{equation}
A simple proof of this relation uses differentiation with respect to $x$,
and Eqs.(\ref{a(x)}, \ref{b(x)}). So, if the denominator $%
1+a(x_{0})\cdot b(x_{0})$ of the quasiclassical propagator Eq.(\ref
{riccpar}), considered as a function of energy $E$, displays a \emph{simple}
zero at $E=E_{b}$, this zero, $E_{b}$, has a natural interpretation as a 
\emph{bound} state energy, provided there exists a finite residue of the retarded propagator at $E=E_{b}+i0^{+}$.

Since it is almost never possible to solve the equations of superconductivity exactly, one needs numerical methods.
The alternative to a straightforward (but costly) numerical solution of the BdG-eigenvalue problem is
the numerical solution of the Eilenberger equations, provided the fundamental condition
of quasi classical theory, $k_{F}\cdot \xi \gg 1$, is valid.
For example, using the quasiclassical approach of Eilenberger, it is comparatively easy to
determine the deep lying bound states $E_{b}$ of localised vortex core fermions (attached to a single
vortex line) as a function of the impact parameters:$E_{b}=E_{b}(y,z)$. 
The summation over (exact)  eigenenergies of the bound states obtained from solving the BdG-eigenvalue problem
 (see Ref.\cite{Caroli}) for a single vortex line becomes equivalent to integrating the quasiclassical propagator with respect
 to the impact parameter.
In certain materials, for example cuprates, the Cooper pairs display (perhaps)
an unconventional $d_{x^{2}-y^{2}}$-symmetry. Results of a quasiclassical
calculation of the bound state spectrum of quasiparticles around a single flux line
in a superconductor with $d_{x^{2}-y^{2}}$-pairing symmetry are published in Refs.\cite{schopohl and maki},\cite{Ichioka}.

For superfluid $^{3}He-B$, a prominent
system with unconventional $p$-wave pairing symmetry, our method can be extended
to the $4\times 4$-Eilenberger propagator for triplet pairing. A calculation of the spectrum of  vortex core fermions 
around the $o$-vortex, the $v$-vortex and also the double core vortex is discussed in Refs. \cite{volovik},\cite{schopohl2}.

We conclude that the quasiclassical propagator may be determined 
solving an initial value problem for a scalar differential equation of the Riccati type.
For numerical calculations this method to solve the Eilenberger equations may be recomended
for its intrinsic stability and speed. Also the Eilenberger approach is a suitable one for parallel computers.

\section{Connection to Linearized Bogoliubov-de\ Gennes Equations}

There exists an interesting connection between the solutions $u(x)$ and $v(x)$ to the \emph{linearized}
BdG-equations (often refered to as Andreev equations \cite{Andreev}), and the solutions
 to the Riccati equations. Along a
characteristic line ${\mathbf r}(x)$ orientated parallel to the Fermi
velocity we have: 
\begin{equation}
-i\hbar v_{F}\frac{\partial }{\partial x}\left[ 
\begin{array}{c}
u(x) \\ 
v(x)
\end{array}
\right] =\left[ 
\begin{array}{cc}
i\tilde{\varepsilon}_{n}(x) & -\Delta (x) \\ 
\Delta ^{\dagger }(x) & -i\tilde{\varepsilon}_{n}(x)
\end{array}
\right] \left[ 
\begin{array}{c}
u(x) \\ 
v(x)
\end{array}
\right]  \label{Andreev}
\end{equation}
We observe that $a(x)$ , the solution to the Riccati Equation Eq.(\ref
{riccatia}) , may be represented as the ratio: 
\begin{equation}
a(x)=i\cdot \frac{u(x)}{v(x)}
\label{u/v}
\end{equation}
This connection between $a(x)$ and the amplitudes $u(x)$ and $v(x)$ is
readily demonstrated: 
\begin{eqnarray}
-\hbar v_{F}\frac{\partial }{\partial x}a(x) &=&\frac{v(x)\left[ -i\hbar
v_{F}\frac{\partial }{\partial x}u(x)\right] -u(x)\left[ -i\hbar v_{F}\frac{%
\partial }{\partial x}v(x)\right] }{\left[ v(x)\right] ^{2}} \\
&=&\frac{v(x)\left[ i\tilde{\varepsilon}_{n}(x)u(x)-\Delta (x)v(x)\right]
-u(x)\left[ \Delta ^{\dagger }(x)u(x)-i\tilde{\varepsilon}_{n}(x)v(x)\right] 
}{\left[ v(x)\right] ^{2}}  \nonumber \\
&=&\left[ 2\tilde{\varepsilon}_{n}\,+\Delta ^{\dagger }(x)\,\cdot
a(x)\right] \,\cdot a(x)-\Delta (x)  \nonumber
\end{eqnarray}
The conclusion is, that any solution to the Eilenberger equations 
(for $i\tilde{\varepsilon}_{n}(x)$ in the upper half of the complex plane for $x\rightarrow -\infty $ )
may be reconstructed from a solution to
the \emph{\ linearized} BdG-differential equations, provided the initial values for $u(x)$ and $v(x)$ are choosen in accordance with the known
asymptotic behaviour of $a(x)$ for $x\rightarrow -\infty $ (see Eq.(\ref
{a(-00)})).

Another interesting feature follows making a polar decomposition of the pair potential:
\begin{equation}
\Delta (x)=\left| \Delta (x)\right| e^{i\phi (x)}
\end{equation}
and making the transformation 
\begin{equation}
a(x)=e^{i\eta (x)}
\end{equation}
in Eq.(\ref{riccatia}). It is readily shown that 
\begin{equation}
v_F\frac \partial {\partial x}\eta (x)-2i\widetilde{\varepsilon }%
_n(x)+2\left| \Delta (x)\right| \sin \left[ \eta (x)-\phi (x)\right] =0
\end{equation}
It is interesting that this differential equation, which is equivalent to
Eq.(\ref{riccatia}), was already derived by J. Bardeen et al. \cite{Bardeen}
in their so called WKBJ-approach to the BdG-equations.

So, the 'new' result is not the Riccati equation Eq.(\ref{riccatia}) itself
(or any equivalent representation). New is the fact, that the full
quasiclassical propagator $\hat{g}\left( {\mathbf r},\theta ,i\varepsilon
_{n}\right) $ (in equilibrium) is simply a rational function (see Eq.(\ref
{riccpar})) of the solutions to a scalar Riccati equation. In turn, solutions to these Riccati equations are 
related by Eq.(\ref{u/v})
 to the solutions of an initial value problem for the linearised Bogoiliubov-de Gennes equations.
This implies, that the standard procedure of wave mechanics to calculate the observables
of superconductivity, namely first solving the Bogolubov-de Gennes eigenvalue
problem and afterwards doing a summation over the eigenfunctions and
eigenvalues to calculate the Green's function, is in deed obsolete and may
be replaced by solving a scalar Riccati equation, or equivalently,
 by solving an initial value problem for 
the linearised BdG-equations, provided the
quasiclassical condition $k_{F}\xi \gg 1$ is fulfilled.

\section{Exact Solutions}

In some cases we may calculate exact solutions to the Eilenberger
equations for a given profile of the pair potential.   First the term ${\mathbf v}_{F}{\mathbf \cdot A}(x)$ in the
definition of $\tilde{\varepsilon}_{n}(x)$ ( local Doppler shift) may be
removed from the diagonal of Eqs.(\ref{Andreev}) making a gauge
transformation, i.e. we may always assume $\tilde{\varepsilon}%
_{n}(x)\rightarrow \varepsilon _{n}$ independent on $x$ . Then, after a
decomposition of the (transformed) pair potential into real and imaginary parts, 
\begin{equation}
\Delta (x)=\Delta _{1}(x)+i\Delta _{2}(x)
\end{equation}
the linearised BdG-equations (\ref{Andreev}) may be given
the form:

\begin{equation}
\left[ \widehat{\tau }_{3}\cdot i\hbar v_{F}\frac{\partial }{\partial x}%
+\Delta _{1}(x)\cdot \widehat{\tau }_{1}-\Delta _{2}(x)\cdot \widehat{\tau }%
_{2}+i\varepsilon _{n}\cdot \widehat{1}\right] \widehat{\tau }_{3}\widehat{%
\psi }(x)=\widehat{0}
\end{equation}
where 
\begin{equation}
\widehat{\psi }(x)=\left( 
\begin{array}{c}
u(x) \\ 
v(x)
\end{array}
\right)
\end{equation}
Introducing an auxiliary spinor $\widehat{\chi }(x)\ $via 
\begin{equation}
\widehat{\tau }_{3}\widehat{\psi }(x)=\left[ \widehat{\tau }_{3}\cdot i\hbar
v_{F}\frac{\partial }{\partial x}+\Delta _{1}(x)\cdot \widehat{\tau }%
_{1}-\Delta _{2}(x)\cdot \widehat{\tau }_{2}-i\varepsilon _{n}\cdot \widehat{%
1}\right] \,\widehat{\chi }(x)  \label{auxilary}
\end{equation}
the following $2^{nd}$ order linear differential equation be derived: 
\begin{eqnarray}
&&\left[ \left( -\hbar ^{2}v_{F}^{2}\frac{\partial ^{2}}{\partial x^{2}}%
+\Delta _{1}^{2}(x)+\Delta _{2}^{2}(x)+\varepsilon _{n}^{2}\right) \cdot 
\widehat{1} \right. \nonumber\\
&& \left. -\hbar v_{F}\left( \frac{\partial \Delta _{1}(x)}{\partial x}%
\cdot \widehat{\tau }_{2}+\frac{\partial \Delta _{2}(x)}{\partial x}\cdot 
\widehat{\tau }_{1}\right) \right] \,\widehat{\chi }(x)=\widehat{0}
\label{spinor}
\end{eqnarray}
Provided an exact solution (in accordance with the initial
conditions Eqs.(\ref{a(-00)}), (\ref{b(+00)})) to
Eq.(\ref{spinor}) can be found, the quasiclassical propagator
may be calculated exactly from Eqs.(\ref{u/v},\ref{riccpar}). This
procedure is a useful principle for the construction of \emph{exact}
solutions to the Eilenberger equations.

The major obstacle to decompose Eq.(\ref{spinor}) just into two
\emph{decoupled} 
scalar differential equations is the spatial \emph{dephasing} between
imaginary and real parts of the gradient of the pair potential, $\frac{%
\partial \Delta _{1}(x)}{\partial x}$ and $\frac{\partial \Delta _{2}(x)}{%
\partial x}$, in Eq.(\ref{spinor}). A simple case occurs, for example, if $\frac{%
\partial \Delta _{2}(x)}{\partial x}=0$ . In this case a rotation around the 
$\widehat{\tau }_{1}$-axis by an angle $\frac{\pi }{2}$ leads to a \emph{%
diagonal} matrix, i.e. the problem may be effectively decoupled into two
scalar differential equations of $2^{nd}$ -order. The well known WKB method%
\cite{WKB} in its standard guise may then be applied to construct an
(approximate) analytical solution in this case. 
If there is no dephasing between real and imaginary parts of the pair potential, some pair
potentials $\Delta (x)$ with model character allow the construction of exact solutions, as we show below.

On the other hand, if there exists dephasing between real and imaginary parts of the pair potential, 
the problem is more difficult. One way to proceed is stratification. This means one approximates the
 pair potential, $\Delta (x)=$ $\Delta_{1}(x)+i\cdot \Delta _{2}(x)$, 
by a sequence of \emph{strata} along $x$, such that $\Delta
_{1}(x)$ and $\Delta _{2}(x)$ become continous and \emph{piecewise linear} 
functions of $x$: 
\begin{eqnarray}
\Delta _{1}(x) &=&c_{1}x+d_{1} \\
\Delta _{2}(x) &=&c_{2}x+d_{2}
\end{eqnarray}
The real constants $d_{1},d_{2},c_{1},c_{2}$ may change from one stratum to another
stratum, like a (linear) spline function. Inside a fixed stratum the $2\times 2$ matrix 
\begin{equation}
\frac{\partial \Delta _{1}(x)}{\partial x}\cdot \widehat{\tau }_{2}+\frac{%
\partial \Delta _{2}(x)}{\partial x}\cdot \widehat{\tau }_{1}=c_{1}\widehat{%
\tau }_{2}+c_{2}\widehat{\tau }_{1}
\end{equation}
may be diagonalised using a suitable,  $x$-\emph{independent} unitary transformation
matrix $\widehat{U}$ of the form 
\begin{eqnarray}
\widehat{U} &=&\frac{1}{\sqrt{2}}\left[ \widehat{\tau }_{3}+\frac{c_{1}}{%
\sqrt{c_{1}^{2}+c_{2}^{2}}}\cdot \widehat{\tau }_{2}+\frac{c_{2}}{\sqrt{%
c_{1}^{2}+c_{2}^{2}}}\cdot \widehat{\tau }_{1}\right] \\
\widehat{U}^{2} &=&\widehat{1} \\
c_{1}\widehat{\tau }_{2}+c_{2}\widehat{\tau }_{1} &=&\widehat{U}\cdot \sqrt{%
c_{1}^{2}+c_{2}^{2}}\,\,\widehat{\tau }_{3}\cdot \widehat{U}
\end{eqnarray}
Since the constants $c_{1},c_{2}$  may change from a given stratum to the neighbouring stratum, the unitary transformation
matrix $\widehat{U}$ changes accordingly. 
Inside a fixed stratum  Eqs.(\ref{spinor}) may be decomposed into two decoupled \emph{scalar}
differential equations\emph{\ }of 2$^{nd}$ -order for the components of the new
spinor: 
\begin{equation}
\widehat{U}\cdot \widehat{\chi }(x)\equiv \widehat{\Phi }(x)=\left( 
\begin{array}{l}
\Phi _{1}(x) \\ 
\Phi _{2}(x)
\end{array}
\right)
\end{equation}
Then the transformed equations assume the familiar form of a
'shifted oscillator': 
\begin{eqnarray}
&&\left[ \left( -\hbar ^{2}v_{F}^{2}\frac{\partial ^{2}}{\partial x^{2}}%
+\left( c_{1}x+d_{1}\right) ^{2}+\left( c_{2}x+d_{2}\right) ^{2}+\varepsilon
_{n}^{2}\right) \cdot \widehat{1} \right. \nonumber \\
&& \left. -\hbar v_{F}\sqrt{c_{1}^{2}+c_{2}^{2}}\,\,%
\widehat{\tau }_{3}\right] \,\widehat{\Phi }(x)=\widehat{0}
\end{eqnarray}
Exact solutions (inside a choosen stratum) may be presented to these differential
equations using the linearly independent parabolic cylinder functions $%
D_{\,\nu }(x)$ and $D_{-\nu -1}(ix)$. The latter special functions are solutions to the differential equation of the harmonic oscillator: 
\begin{equation}
\left( \frac{\partial ^{2}}{\partial x^{2}}+\nu +\frac{1}{2}-x^{2}\right)
D(x)=0
\end{equation}
The components of $\widehat{\Phi }(x)$ are related to $\widehat{\chi }(x)$
via the relation $\widehat{\chi }(x)=\widehat{U}\cdot \widehat{\Phi }(x)$ ,
i.e. the changes of the constants $c_{1}$ and $c_{2}$ from one
stratum to the next stratum determine also the admixture of $\Phi _{1}(x)$
and $\Phi _{2}(x)$ , and henceforth the spinor $\widehat{\chi }(x)$: 
\begin{eqnarray}
\chi _{1}(x) &=&\frac{1}{\sqrt{2}}\left[ \Phi _{1}(x)+\frac{c_{2}-ic_{1}}{%
\sqrt{c_{1}^{2}+c_{2}^{2}}}\Phi _{2}(x)\right] \\
\chi _{2}(x) &=&\frac{1}{\sqrt{2}}\left[ \Phi _{2}(x)+\frac{c_{2}+ic_{1}}{%
\sqrt{c_{1}^{2}+c_{2}^{2}}}\Phi _{1}(x)\right]
\end{eqnarray}
Finally, from Eq.(\ref{auxilary}) the ratio $a(x)=i\cdot \frac{u(x)}{v(x)}$ may be
calculated such that $a(x)$ remains a smooth function of $x$ . Once $b(x)$ has been found from a similar consideration
(or from $a(x)$ by a symmetry argument) the quasiclassical propagator follows from Eq.(\ref{riccpar}).

Closing this section we give three examples of pair potential profiles $%
\Delta (x)$ that allow an exact solution to the Eilenberger equations.
Allthough these pair potential are not selfconsistent they are helpful for
a qualitative physical understanding:

\begin{eqnarray}
\Delta (x) &=&\Delta _\infty \cdot \frac{x+iy}\xi  \label{m1} \\
\Delta (x) &=&\Delta _\infty \cdot \frac{x+iy}{\sqrt{x^2+y^2}}  \label{m2} \\
\Delta (x) &=&\Delta _\infty \cdot \tanh (\frac x\xi )  \label{m3}
\end{eqnarray}

The model Eq.(\ref{m1}) represents the \emph{inner} core of a vortex, and an
exact solution to the Eilenberger equations may be constructed in terms of
the parabolic cylinder functions along the lines explained above\cite
{schopohl2}.

The model Eq. (\ref{m2}) represents the \emph{outer} core of a vortex ( it
describes a pure phase vortex), and may actually be solved exactly for the
special case $i\varepsilon _{n}\rightarrow E=\left| \Delta _{\infty }\right| 
$. 
\begin{eqnarray}
a(x) &=&\frac{\left( 1-2iW\right) \left( y+\sqrt{x^{2}+y^{2}}\right)
+(1+2iW)\cdot ix}{\left( 1+2iW\right) \left( y+\sqrt{x^{2}+y^{2}}\right)
-(1-2iW)\cdot ix} \\
W &=&\sqrt{\frac{1}{4}-\frac{y}{\xi }} \\
\xi  &=&\frac{\hbar v_{F}}{\left| \Delta _{\infty }\right| }
\end{eqnarray}
At the gap edge, $E=\left| \Delta _{\infty }\right| $ , the corresponding
expression for the quasiclassical propagator, Eq. (\ref{riccpar}), displays 
\emph{algebraic} decay for $x\rightarrow \pm \infty $. Certainly, it would
be nice to know the exact solution also for other energies $E$ , but this
problem is still unsolved. Nevertheless, a qualitatively correct (but not
exact) solution to Eq.(\ref{spinor}) for the case of a vortex line may be
found \cite{schopohl2} using a method of Stueckelberg\cite{Stueckelberg}.

The third model, Eq.(\ref{m3}), is simpler than the previous model, Eq.(\ref
{m2}), because the pair potential $\Delta (x)$ is a real function (no
dephasing of components of spinor $\widehat{\chi }$). The differential
equation for $\widehat{\chi }(x)$, Eq.(\ref{spinor}), may be solved exactly
in terms of the hypergeometric function\cite{Morse-Feshbach}). 
A particularly simple looking analytic solution results if the size parameter $%
\xi $ of the domain wall described by Eq.(\ref{m3}) assumes the special value
 $\xi =\frac{\hbar v_{F}}{\Delta _{\infty }}$: 
\begin{eqnarray}
a(x) &=&\frac{\varepsilon _{n}-\sqrt{\varepsilon _{n}^{2}+\Delta _{\infty
}^{2}}+\Delta _{\infty }\tanh \frac{x}{\xi }}{\varepsilon _{n}-\sqrt{%
\varepsilon _{n}^{2}+\Delta _{\infty }^{2}}-\Delta _{\infty }\tanh \frac{x}{%
\xi }} \\[0.5in]
\widehat{g}(x) &=&\frac{-i\pi }{\sqrt{\varepsilon _{n}^{2}+\Delta _{\infty
}^{2}}}\cdot \left[ 
\begin{array}{c}
\left( \varepsilon _{n}+\frac{\Delta _{\infty }^{2}}{2\varepsilon _{n}}\frac{%
1}{\cosh ^{2}\frac{x}{\xi }}\right) \cdot \widehat{\tau }_{3} \\ 
-\;\Delta _{\infty }\tanh \frac{x}{\xi }\cdot \widehat{\tau }_{2} \\ 
-\;i\frac{\Delta _{\infty }^{2}}{2\varepsilon _{n}}\frac{1}{\cosh ^{2}\frac{x%
}{\xi }}\cdot \widehat{\tau }_{1}
\end{array}
\right]
\end{eqnarray}
We see that the explicit analytical expression for the quasiclassical
propagator, $\widehat{g}(x)$, nicely reveals the existence of a mid gap
state (a quasiparticle bound state with an excitation energy $E$ around zero), assuming
 we make the analytical continuation $i\varepsilon _{n}\rightarrow E+i0^{+}$. 
The existence of such a mid gap state is in agreement with our previous remarks
associated with Eq.(\ref{bound state condition}). 

\section{Outlook}

In this article we have shown how to parametrise the $2\times 2$-Eilenberger
equations in terms of the solutions to a scalar Riccati equation, or equivalently,
in terms of the solutions to an initial value problem for the linearised BdG-equations. 
The \emph{latter} method for the reconstruction of the Eilenberger propagator may be extended \cite
{schopohl2} to the full $4\times 4$-matrix equations of the quasiclassical
theory of superconductivity and superfluidity (including paramagnetic effects). 
For a generalisation of the method to non equilibrium see \cite{Zaikin},\cite{Eschrig}). 
For an authoritative review of the
quasiclassical theory see Ref.\cite{Rainer}.

\medskip
\textbf{Acknowledgments}: It is a pleasure to thank D. Rainer and J.A. Sauls
for discussions and hospitality during the 'First Workshop on Quasiclassical
Methods in Superconductivity and Superfluidity', at Verditz, Carinthia, Austria, during
March 11-16, 1996 (proceedings to be published in \cite{Springer}). I thank
very much K. Nagai for discussions and for bringing Refs.\cite{Nagai-I},\cite
{Nagai} to my attention. Last not least I am grateful to K. Maki for the
interest he has taken in this work and for encouragement.


\end{document}